Intra-Day Seasonality in Foreign Exchange Market Transactions

John Cotter and Kevin Dowd[*]

Abstract
This paper examines the intra-day seasonality of transacted limit and market orders in the DEM/USD foreign exchange market. Empirical analysis of completed transactions data based on the Dealing 2000-2 electronic inter-dealer broking system indicates significant evidence of intraday seasonality in returns and return volatilities under usual market conditions. Moreover, analysis of realised tail outcomes supports seasonality for extraordinary market conditions across the trading day.

[*] John Cotter is at the Centre for Financial Markets, School of Business, University College Dublin, Carysfort Avenue, Blackrock, Co. Dublin, Ireland; email: john.cotter@ucd.ie. Kevin Dowd is at the Centre for Risk and Insurance Studies, Nottingham University Business School, Jubilee Campus, Nottingham NG8 1BB, UK; email: Kevin.Dowd@nottingham.ac.uk. The authors thank Charles Goodhart, Jon Danielsson and Richard Payne for fruitful conversations on the subject, but remain responsible for any remaining errors. Cotter's contribution to the study has been supported by a University College Dublin School of Business research grant.

## 1. Introduction:

In recent times much has been made of the trading revolution in currency markets brought about through the screen-based electronic trading and broking systems that have come to dominate foreign exchange trading activities. The importance of these systems is illustrated by the fact that, according to the most recent Bank for International Settlements (BIS) survey on foreign exchange, electronic trading makes up 48% of activity in the largest market, the UK (Williams, 2005).

This study examines the intra-day seasonalities of returns and volatilities in FX transactions. More specifically, it examines the intra-day return and volatility seasonalities for limit and market orders involving the DEM/USD exchange rate from the D2000–2 electronic FX broking system. The analysis is based on actual transaction data rather than the more common (but less reliable) use of indicative quotes in which there is no firm commitment to transact on the stated terms. Limit orders represent an order to buy or sell at some prespecified price, whereas market orders are orders for immediate execution at whatever price can be obtained.

Previous microstructure studies have demonstrated the importance of order type return and volatility characteristics. For example, Hasbrouck and Saar (2004) find evidence of a market order certainty effect where increased limit order volatility is associated with a reduction in the proportion of limit orders in incoming order flow.[1] Relatedly, Hasbrouck and Harris (1996) compare the execution performance of market and limit order and show that limit order trading strategies generally perform best across various spreads, order sizes and position (buy or sell).

---

[1] The market order certainty effect implies that risk-averse traders place a premium on a definite outcome. Moreover, higher volatility increases the dispersion of wealth outcomes for a limit order strategy thereby making market orders more attractive.

There is also reason to believe that traders will alter their trading strategies and associated order mechanisms during periods of extreme market movements (Goldstein and Kavajecz (2004)). We therefore believe that it is also important to distinguish between normal and extraordinary market conditions (Longin, 2000). To do so, we also examine the intra-day seasonality of the tail behaviour of market and limit orders to determine if seasonalities and trading patterns differ between normal and extreme market conditions.

We model this tail behaviour using Extreme Value Theory (EVT). Many previous studies have documented the presence of heavy-tails for the distribution of exchange rate price changes over different frequencies and under different institutional frameworks (e.g., Cotter, 2005). Less is known about the features of exchange rate returns according to order type and this paper investigates seasonality for the extreme returns of market and limit orders.

The rest of the paper proceeds as follows. Section 2 outlines the foreign exchange market and limit and market order features of the D2000-2 dataset on which the study is based. Section 3 presents Extreme Value Theory and explains its properties and the measurement techniques use to examine market and limit order tail behaviour in the market considered. Section 4 presents the empirical findings. Some conclusions are given in section 5.

**2. Data Considerations**

2.1 Foreign exchange market

The foreign exchange market is a highly decentralized market. Market participants seldom physically meet, but rather operate in separate offices of the major commercial banks. Trading typically takes place using telephone and electronic means. There is no central regulator governing these trading relationships, although private regulation exists to ensure a code of conduct in market transactions. Traditionally, individual dealers had only to disclose information to the trade's counterparty and no mechanism existed to observe all market activity. However, with the introduction of electronic broker dealer trades, foreign exchange traders have access to information on all trading activity and can now assess ongoing market conditions in real-time.

The spot foreign exchange market is the largest spot asset market in the world and the most actively traded exchange rate over the period of this study – October 6, 1997 through October 11, 1997 - was the DEM/USD exchange rate. To give an illustration, the average daily volume for all currency trading was US$568 billion, an amount which dwarves the average US$75 billion traded daily in the New York Stock Exchange, and transactions involving the DEM/USD exchange rate accounted for 20% of total trading (BIS, 2004).

2.2 D2000-2 data set

The limit and market order foreign exchange data employed in this study are from the D2000–2 electronic FX inter-dealer broking system run by Reuters. This unique brokerage system is one of two main electronic brokers in this market, the other operated by the EBS partnership. Unlike other systems which deal only with single dealers, D2000-2 allows for an examination of the activities of multiple traders (see, e.g., Evans and Lyons (1999)). As the system covers multiple trader activity, it provides for a comprehensive description of market order information. Furthermore, because it refers to actual transaction data, use of the D2000-2 data set overcomes problems associated with the use of indicative quotes which can be unreliable because they do not refer to binding trade commitments.[2]

Foreign exchange trading can be distinguished under three headings: customer dealer trades, direct inter-dealer trades and brokered dealer trades. Of these the majority of foreign exchange trading are inter-dealer based and were traditionally completed by telephone. However, more recently, the introduction of electronic broking systems has resulted in a major change in how dealers trade and has led to rapid growth in the numbers of brokered trades completed electronically. Estimates for the two main electronic brokerage systems suggest that over the October 1997 trading week covered in this study approximately 40% of all trade in London was completed by EBS and D2000-2. The recent 2004 BIS survey indicates that this proportion has since risen to 66% (Williams, 2005).

---

[2] Further shortcomings in applying these indicative quotes in microstructure studies include the lack of traded volume information and the absence of details on the timescale of quotes (Danielsson and Payne, 2002).

The D2000-2 data set contains all trading activity in USD/DEM for the trading week covering the 6th to the 10th of October 1997, incorporating 130,535 entries in all. Of these, over 100,000 transactions took the form of limit orders. The information stored for analysis within the data set is comprehensive.[3] Of particular concern to this study are the following limit and market order quotes: limit buy quote, limit sell quote, market buy quote and market sell quote.[4] The average size of a filled limit order is $2m whereas for market orders it is $3m. Thus, limit orders are a more popular and more liquid type of trade, but limit orders are also typically smaller.

Importantly, D2000–2 operates as a pure limit order market governed by rules of price and time priority. Limit orders queue so that market orders hit the best outstanding limit order on a given side of the market. This gives the D2000-2 user the best limit buy and sell prices, plus quantities available at these prices and a record of recent transaction activity. Only filled orders are analysed, and limit orders get filled one time in three whereas market orders are always partially or fully filled. This study obtains midquotes using the best bid and offer quotes at the end of each trading interval measured in calendar time. To convert to calendar time, it uses a sampling frequency of 20 seconds. This choice of frequency ensures that the information usage inherent in the data set is maximized and avoids omitting observations during periods of intense activity, but also helps to ensure that intervals do not suffer from a thin trading bias.

---

[3] For each trading entry, there are ten fields of information detailing the type of event to which it refers, timestamps within one hundredth of a second, plus price and quantities.

[4] Furthermore, the information on limit orders contains details of timestamps for entry and exit times, a buy/sell indicator, quantity available, quantity traded and price; whereas for market orders, the information contains details of quantity transacted, price, a timestamp and whether the trade is buyer or seller initiated.

The majority of trading activity takes place between 9-18 GMT when almost 70% of all orders are processed and the remaining intervals involve relatively thin trades. The quietest period is between 21-24 GMT when less than 1% of orders are filled. Following Andersen et alia (2001), all findings are presented for 5-minute intervals to minimize the serial correlation in the bid-ask spread returns induced by non-synchronous trading. . Returns are calculated as the first difference of log prices and volatility is proxied by absolute returns.

## 3. Theoretical framework

An appropriate theoretical framework to examine the limit and market order tail behaviour is provided by Extreme value theory (EVT).[5] EVT distinguishes three types of asymptotic distributions, Gumbel, Weibull and the one of concern to this study, the heavy-tailed Fréchet distribution. EVT models only tail returns, unlike other fully parametric approaches (e.g.. one based on elliptical distributions) that model the full distribution of realised returns.

To begin, we assume that the random return variable R is independent and identically distributed (iid) and belongs to the true unknown cumulative probability density function $F(r)$.[6] Let $(M_n)$ be the maxima of $n$ random variables such that $M_n = max \{R_1, R_2,..., R_n$.

---

[5] Only salient features are presented with comprehensive details in Embrechts et al (1997).
[6] GARCH modelling of financial returns invalidates this iid assumption. However, de Haan et al (1989) provide extensions for processes with non-iid realisations that match the characteristics of financial returns.

The probability that the maximum value $M_n$ exceeds a certain price change, $r$,[7] is then given by

$$P\{M_n > r\} = P\{R_1 > r, ..., R_n > r\} = 1 - F^n(r) \qquad (1)$$

Whilst the exact distribution of $F^n(r)$ is unknown, the regular variation at infinity property gives a necessary and sufficient condition for $F^n(r)$ to converge to the heavy-tailed extreme value distribution. This condition is as follows

$$\lim_{t \to \infty} \frac{1 - F(tr)}{1 - F(t)} = r^{-\alpha} \qquad (2)$$

This condition implies that the distribution $F(r)$ obeys a Pareto distribution. The Pareto distribution allows heavy-tailed distributions to exhibit identical behaviour after asymptotic convergence.

Assuming the distribution exhibits the regular variation at infinity property then a first order Taylor expansion gives the following asymptotically:

$$1 - F^n(r) \approx ar^{-\alpha} \qquad (3)$$

where $a$ represents the scaling constant and $\alpha$ is the tail index, for $\alpha > 0$ and for $r \to \infty$. The tail index, $\alpha$, measures the degree of tail thickness. By l'Hopital's rule the Student-t, and symmetric non-normal sum-stable distributions, and certain ARCH processes with an unconditional stationary distribution all exhibit the property of regular variation at infinity property, and this means that their asymptotic tails decline by a power function.

---

[7] Adjustments for lower tail and common tail statistics are easily computed although the theoretical framework is presented for upper tail statistics following convention.

These are therefore heavy-tailed and have a low tail index. More importantly, all these distributions exhibit identical behaviour far out in the tails.

The tail index, $\alpha$, also measures the number of bounded moments, and financial studies have commonly cited $\alpha$ values between 2 and 4 suggesting that higher moments of the returns series may not exist (e.g., Loretan and Phillips, 1994).

We estimate the tail index using the Hill (1975) moment estimator. The Hill estimator represents a maximum likelihood estimator of the inverse of the tail index:

$$\gamma(m) = 1/\alpha = (1/m) \sum [log\ r_{(n+1-i)} - log\ r_{(n-m)}] \qquad for\ i = 1....m \qquad (5)$$

This tail estimator is predicated on a tail threshold $m$, and is known to be asymptotically normal (Goldie and Smith, 1987).

However, an unresolved issue in tail estimation is to determine the optimal tail threshold, $m$. The Hill estimator is asymptotically unbiased but suffers from small sample bias which limits its application to many financial return data sets. The estimation problem is due to a trade-off between the bias and variance of the estimator with the bias (variance) decreasing (increasing) with the number of tail values used. To mitigate its potential small-sample bias, we use a weighted least squares regression due to Huisman et al (2001). Here, a number of different Hill tail estimates are measured and the following regression is estimated:

$$\gamma(m) = \gamma + \beta m + \varepsilon(m) \qquad for\ m = 1,....,\eta \qquad (6)$$

The estimate of γ using the weighted least squares regression is expected to be an unbiased estimator of the tail estimator.

**4. Empirical Findings:**

We begin by examining the stylized facts of the returns and volatility series for market and limit orders across the trading day. Summary statistics are given in Table 1 with related time series plots in Figure 1. In general, we find the conventional stylized facts of excess (positive) skewness and excess kurtosis (and therefore non-normality) for both order types regardless of trading interval. The heavy tails exhibited by the kurtosis statistic are more pronounced during the thin trading periods both early and late in the day. For ease of comparison returns and volatility estimates are presented on an annualised basis. The average returns and standard deviation, whilst similar in magnitude, show that limit order return realisations are greater than their market order counterparts.

INSERT TABLE 1 HERE

INSERT FIGURE 1 HERE

Both limit and market orders show intra-day seasonality. However, there are also some noticeable differences between limit and market orders and in their intraday seasonality behaviour. Figure 1 clearly shows a daily pattern emerging for the market order returns and volatilitiess. The average dispersion of the volatility proxies also suggests seasonality across a day with limit orders showing greater magnitude of realisations than market orders.

We now turn to the relationships for limit and market order returns and their volatility characteristics across the trading day. In Table 2 we present the correlation coefficients for market and limit realisations for 3-hour blocks. Very weak relationships for both return and volatility realisations across the trading day are recorded with most of the correlations close to zero and many negative. The strongest correlation for market order returns occurs for the 0-3 GMT and 18-21 GMT intervals (0.227) and for limit orders it is during the 3-6 GMT and 6-9 GMT intervals (0.123). Similar magnitudes are recorded for correlations of absolute returns.[8] Thus, there are no strong correlations between either returns or volatilities across the trading day.

INSERT TABLE 2 HERE

Table 3 presents estimates of the Hill tail index and its standard errors. These results show large variations of tail behaviour across the trading day. For instance, the point estimates for lower tails range between 1.33 and 5.78 for market orders and between 2.02 and 5.25 for limit orders. Generally the small values are associated with thin trading periods whereas the larger values occur during the more actively traded periods. Figure 2 presents the modified Hill tail estimators by order type and trading interval. The plots reinforce our findings that tail estimates exhibit intraday seasonality. This Figure also shows that the tail values for both limit and market order are generally largest (smallest) during the most actively (thinly) traded GMT interval.

---

[8] We also find the patterns of correlation between market and limit order realisations to be reasonably similar during heavier trading intervals but diverge during thinly traded intervals suggesting seasonality in the relationship between these order types. Results available on request.

INSERT TABLE 3 HERE

INSERT FIGURE 2 HERE

The variation in tail estimates can be further emphasised by assessing the distributional implications of the estimates. First we note that the existence of the heavy tails for limit and market orders is confirmed for all trading periods: this is indicated by the significant t-statistic that rejects the null, $H_o: \gamma = 0$. Values for the hypotheses, $H_o: \gamma = 2$, and $H_o: \gamma = 4$ determine the existence of $2^{nd}$ and $4^{th}$ moments respectively. Similarly, the hypothesis of a defined (i.e., finite) variance is never rejected for all intervals. In contrast, evidence for the existence of the $4^{th}$ moment of the underlying distribution is mixed: slightly less than half of the intervals exhibiting tail values that support the hypothesis. Rejection of the hypothesis for the existence of a $4^{th}$ moment varies throughout the day for order type and across trading position. Moreover, the results in Table 3 suggest that the degree of tail-heaviness is higher (lower) for heavily (thinly) traded intervals. Also, the market order estimates are generally larger (smaller) for these heavily (thinly) traded intervals. And, whilst there are exceptions, the intraday Hill plots suggest that limit orders are less heavily-tailed than market order returns during the standard GMT trading day whereas the converse is true during other periods.

## 5. Conclusion and Summary

This paper has examined risk and return realisations for FX market and limit orders across a trading day. We analyse a unique data set of transacted exchange rates overcoming the common use of indicative quotes that may or may not be binding. We find that both limit and market orders exhibit seasonality through the variation of their returns and volatility across the trading day, and this is so (although in different ways) for both ordinary and extraordinary market conditions  Our work provides some preliminary findings about the behaviour of intra-day seasonalities in FX markets, but much more work will be needed to determine how typical our findings might be of seasonalities in other periods or other markets, and to determine whether, and if so how, seasonalities might be changing over time.

**FIGURES:**

Figure 1: Plots of 5 minute returns and volatility across a trading day

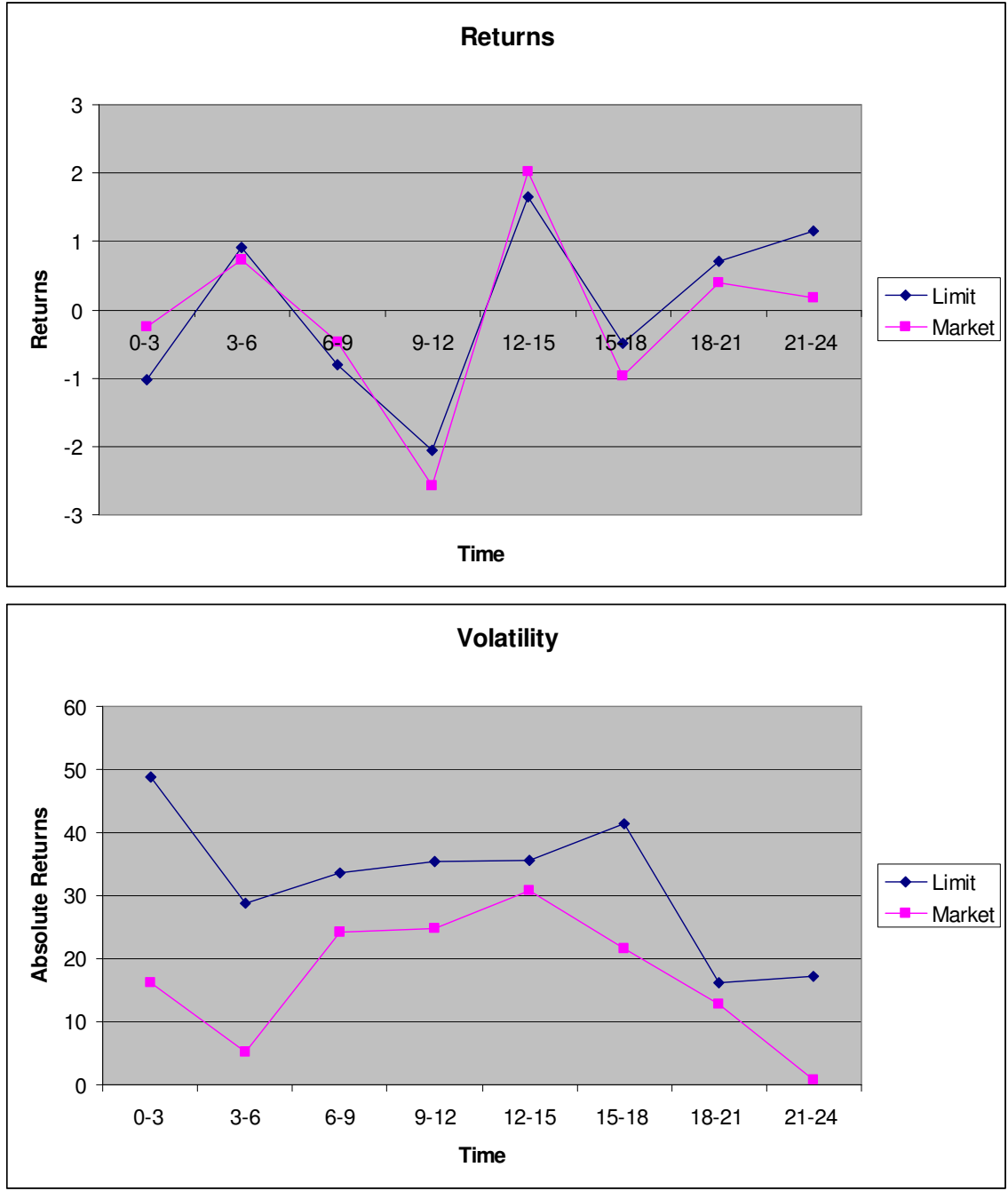

Figure 2: Intraday seasonality plots of tail estimates

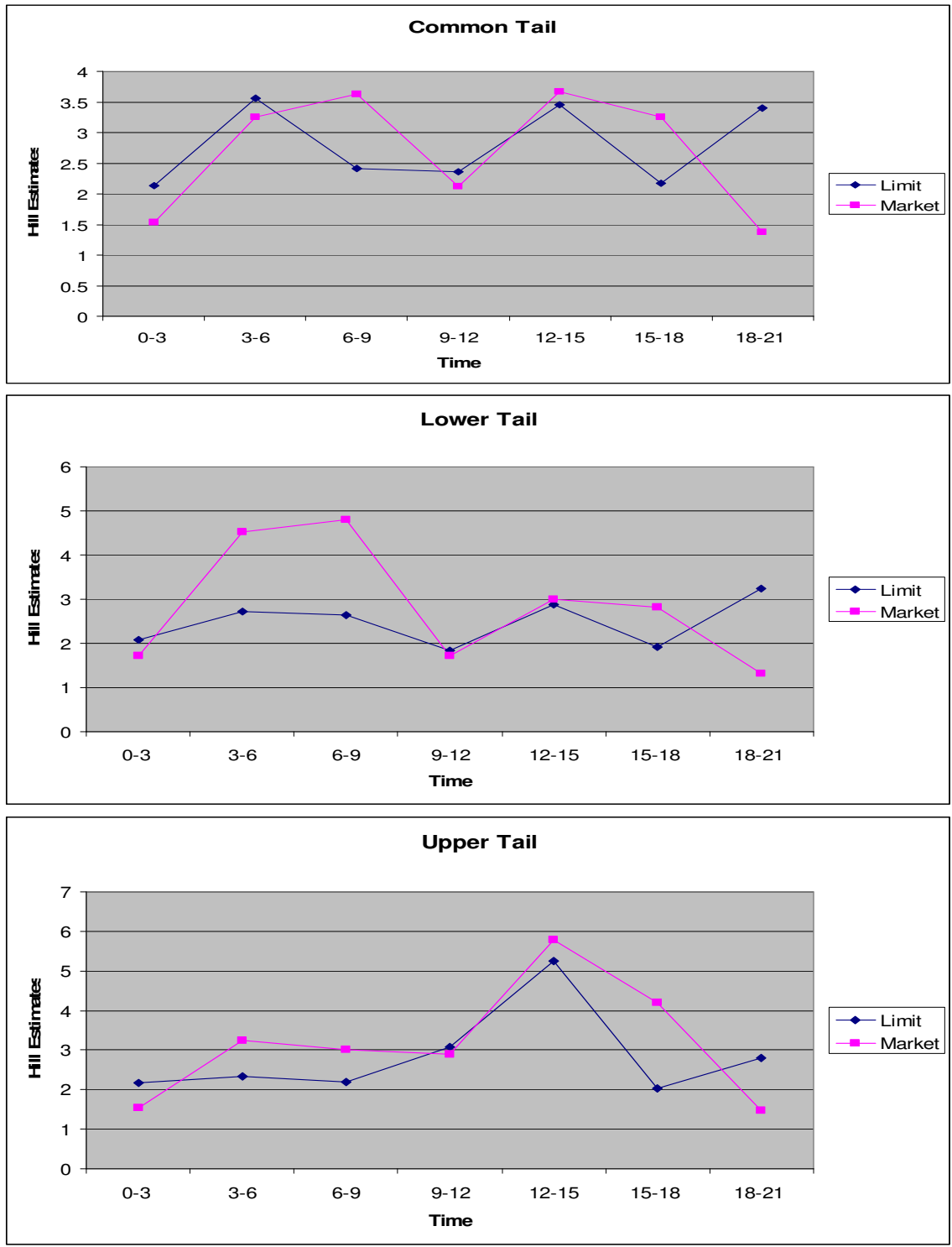

Notes: 21-24GMT is excluded due to a lack of trading activity for market orders.

TABLES:

Table 1: Summary statistics for 5 minute returns and volatility for different intervals

|  | **Limit** | | | | | **Market** | | | | |
| **Returns** | **Mean** | **Std Dev** | **Skew** | **Kurt** | **Norm** | **Mean** | **Std Dev** | **Skew** | **Kurt** | **Norm** |
| 0-3 | -1.033 | 92.851 | 0.234 | 12.100 | 0.187 | -0.241 | 51.218 | 0.341 | 54.900 | 0.307 |
| 3-6 | 0.906 | 56.909 | 0.357 | 6.100 | 0.230 | 0.733 | 12.355 | 1.770 | 9.840 | 0.390 |
| 6-9 | -0.809 | 50.769 | 0.017 | 6.190 | 0.113 | -0.472 | 32.723 | 0.280 | 1.360 | 0.081 |
| 9-12 | -2.059 | 57.583 | 2.400 | 15.900 | 0.132 | -2.568 | 46.426 | 4.200 | 31.200 | 0.179 |
| 12-15 | 1.647 | 47.324 | 0.276 | 1.640 | 0.055 | 2.022 | 41.633 | 0.562 | 2.240 | 0.099 |
| 15-18 | -0.497 | 73.158 | 0.085 | 10.400 | 0.171 | -0.966 | 33.172 | 0.272 | 3.340 | 0.142 |
| 18-21 | 0.706 | 26.358 | 0.329 | 4.940 | 0.149 | 0.403 | 48.822 | 0.382 | 71.600 | 0.304 |
| 21-24 | 1.153 | 43.281 | 1.120 | 14.000 | 0.344 | 0.168 | 5.451 | 7.260 | 85.600 | 0.496 |
| **Volatility** | | | | | | | | | | |
| 0-3 | 48.822 | 78.624 | 3.360 | 15.100 | 0.267 | 16.174 | 48.597 | 7.120 | 58.100 | 0.370 |
| 3-6 | 28.829 | 49.046 | 2.390 | 5.580 | 0.293 | 5.197 | 11.232 | 2.900 | 9.650 | 0.389 |
| 6-9 | 33.546 | 37.964 | 2.830 | 10.700 | 0.210 | 24.261 | 21.865 | 1.510 | 2.880 | 0.150 |
| 9-12 | 35.343 | 45.452 | 4.500 | 28.300 | 0.221 | 24.785 | 39.312 | 6.140 | 47.200 | 0.264 |
| 12-15 | 35.643 | 31.075 | 1.700 | 4.130 | 0.148 | 30.701 | 28.080 | 1.770 | 5.210 | 0.171 |
| 15-18 | 41.484 | 60.129 | 3.360 | 13.600 | 0.245 | 21.565 | 25.160 | 1.900 | 4.540 | 0.196 |
| 18-21 | 16.174 | 20.742 | 2.290 | 6.090 | 0.218 | 12.879 | 47.100 | 8.320 | 73.400 | 0.393 |
| 21-24 | 17.102 | 39.740 | 3.390 | 13.982 | 0.344 | 0.884 | 5.376 | 8.520 | 85.400 | 0.526 |

Notes: With the exception of (skew)ness and (kurt)osis coefficients, all values are expressed in percentage form. The skewness statistic is a measure of distribution asymmetry with symmetric returns having a value of zero. The kurtosis statistic measures the shape of a distribution vis-à-vis a normal distribution with a gaussian density function having a value of 3. (Norm)ality is formally examined with the Kolmogorov-Smirnov test which indicates a gaussian distribution with a value of zero. All skewness, kurtosis and normality coefficients are significant at the 5 percent level.

Table 2: Correlation matrices of returns and volatility across a trading day

| Returns Market | 0-3 | 3-6 | 6-9 | 9-12 | 12-15 | 15-18 | 18-21 | 21-24 |
|---|---|---|---|---|---|---|---|---|
| 0-3 | 1.000 | | | | | | | |
| 3-6 | 0.088 | 1.000 | | | | | | |
| 6-9 | 0.072 | 0.104 | 1.000 | | | | | |
| 9-12 | 0.039 | -0.122 | -0.093 | 1.000 | | | | |
| 12-15 | -0.241 | 0.053 | -0.099 | -0.126 | 1.000 | | | |
| 15-18 | 0.024 | 0.002 | -0.041 | 0.162 | 0.023 | 1.000 | | |
| 18-21 | 0.227 | -0.161 | 0.040 | -0.012 | -0.094 | 0.079 | 1.000 | |
| 21-24 | 0.092 | -0.085 | 0.018 | 0.051 | -0.102 | 0.016 | -0.061 | 1.000 |
| **Limit** | | | | | | | | |
| 0-3 | 1.000 | | | | | | | |
| 3-6 | -0.011 | 1.000 | | | | | | |
| 6-9 | -0.208 | 0.123 | 1.000 | | | | | |
| 9-12 | -0.052 | 0.039 | -0.149 | 1.000 | | | | |
| 12-15 | -0.031 | -0.035 | -0.073 | -0.118 | 1.000 | | | |
| 15-18 | 0.059 | 0.114 | 0.057 | -0.006 | -0.057 | 1.000 | | |
| 18-21 | 0.006 | -0.283 | 0.008 | 0.021 | 0.124 | -0.177 | 1.000 | |
| 21-24 | 0.034 | -0.002 | -0.021 | 0.038 | -0.050 | 0.012 | 0.008 | 1.000 |

| Volatility Market | 0-3 | 3-6 | 6-9 | 9-12 | 12-15 | 15-18 | 18-21 | 21-24 |
|---|---|---|---|---|---|---|---|---|
| 0-3 | 1.000 | | | | | | | |
| 3-6 | 0.132 | 1.000 | | | | | | |
| 6-9 | 0.005 | 0.004 | 1.000 | | | | | |
| 9-12 | -0.028 | -0.013 | -0.005 | 1.000 | | | | |
| 12-15 | 0.218 | -0.027 | 0.030 | 0.100 | 1.000 | | | |
| 15-18 | -0.022 | -0.048 | -0.046 | -0.011 | -0.089 | 1.000 | | |
| 18-21 | 0.143 | -0.007 | -0.042 | -0.047 | 0.122 | -0.014 | 1.000 | |
| 21-24 | 0.038 | -0.044 | -0.108 | -0.004 | -0.119 | 0.070 | 0.260 | 1.000 |
| **Limit** | | | | | | | | |
| 0-3 | 1.000 | | | | | | | |
| 3-6 | -0.058 | 1.000 | | | | | | |
| 6-9 | 0.159 | -0.002 | 1.000 | | | | | |
| 9-12 | -0.049 | -0.051 | 0.036 | 1.000 | | | | |
| 12-15 | -0.119 | 0.054 | 0.024 | -0.042 | 1.000 | | | |
| 15-18 | 0.028 | 0.013 | -0.026 | -0.103 | -0.016 | 1.000 | | |
| 18-21 | -0.040 | 0.237 | -0.057 | -0.068 | 0.179 | 0.149 | 1.000 | |
| 21-24 | -0.018 | -0.036 | -0.118 | -0.043 | -0.080 | -0.014 | -0.029 | 1.000 |

Notes: The off-diagonal values represent correlation between different intervals (eg. 3-6 GMT and 0-3 GMT) for both limit and market order returns and volatility (absolute returns).

Table 3: Tail estimates for 5 minute limit and market order returns for different intervals

| | Limit | | | | Market | | | |
|---|---|---|---|---|---|---|---|---|
| | $\gamma$ | $\gamma 0 =$ | $\gamma 2 =$ | $\gamma 4 =$ | $\gamma$ | $\gamma 0 =$ | $\gamma 2 =$ | $\gamma 4 =$ |
| Common Tail | | | | | | | | |
| 0-3 | 2.14 | 3.52 | 0.23 | -3.06 | 1.53 | 3.49 | -1.07 | -5.64 |
| | -0.31 | | | | -0.22 | | | |
| 3-6 | 3.56 | 5.03 | 2.21 | -0.62 | 3.25 | 6.23 | 2.40 | -1.43 |
| | -0.36 | | | | -0.27 | | | |
| 6-9 | 2.41 | 2.97 | 0.51 | -1.96 | 3.63 | 2.92 | 1.31 | -0.30 |
| | -0.41 | | | | -0.63 | | | |
| 9-12 | 2.36 | 3.03 | 0.47 | -2.10 | 2.12 | 2.45 | 0.14 | -2.18 |
| | -0.40 | | | | -0.44 | | | |
| 12-15 | 3.45 | 2.95 | 1.24 | -0.47 | 3.67 | 2.74 | 1.25 | -0.25 |
| | -0.60 | | | | -0.68 | | | |
| 15-18 | 2.17 | 3.57 | 0.28 | -3.01 | 3.25 | 3.05 | 1.17 | -0.70 |
| | -0.31 | | | | -0.55 | | | |
| 18-21 | 3.40 | 4.42 | 1.81 | -0.79 | 1.38 | 2.32 | -1.03 | -4.39 |
| | -0.39 | | | | -0.30 | | | |
| Lower Tail | | | | | | | | |
| 0-3 | 2.09 | 2.13 | 0.09 | -1.95 | 1.73 | 2.81 | -0.45 | -3.70 |
| | -0.50 | | | | -0.31 | | | |
| 3-6 | 2.72 | 2.03 | 0.54 | -0.96 | 4.53 | 4.79 | 2.67 | 0.56 |
| | -0.68 | | | | -0.48 | | | |
| 6-9 | 2.65 | 2.12 | 0.52 | -1.08 | 4.81 | 1.88 | 1.10 | 0.32 |
| | -0.64 | | | | -1.31 | | | |
| 9-12 | 1.84 | 1.89 | -0.16 | -2.22 | 1.73 | 1.00 | -0.15 | -1.30 |
| | -0.50 | | | | -0.89 | | | |
| 12-15 | 2.89 | 2.32 | 0.71 | -0.89 | 3.01 | 1.93 | 0.65 | -0.64 |
| | -0.64 | | | | -0.79 | | | |
| 15-18 | 1.92 | 2.35 | -0.10 | -2.56 | 2.82 | 1.86 | 0.54 | -0.78 |
| | -0.42 | | | | -0.77 | | | |
| 18-21 | 3.25 | 3.15 | 1.22 | -0.72 | 1.33 | 1.68 | -0.84 | -3.36 |
| | -0.53 | | | | -0.40 | | | |
| Upper Tail | | | | | | | | |
| 0-3 | 2.17 | 2.85 | 0.22 | -2.40 | 1.53 | 2.16 | -0.67 | -3.50 |
| | -0.39 | | | | -0.36 | | | |
| 3-6 | 2.33 | 2.80 | 0.39 | -2.01 | 3.25 | 3.85 | 1.48 | -0.89 |
| | -0.42 | | | | -0.43 | | | |
| 6-9 | 2.20 | 2.29 | 0.21 | -1.87 | 3.02 | 2.28 | 0.77 | -0.74 |
| | -0.49 | | | | -0.67 | | | |
| 9-12 | 3.07 | 1.90 | 0.67 | -0.57 | 2.89 | 1.66 | 0.51 | -0.64 |
| | -0.82 | | | | -0.89 | | | |
| 12-15 | 5.25 | 2.74 | 1.70 | 0.65 | 5.78 | 1.53 | 1.00 | 0.47 |
| | -0.98 | | | | -1.93 | | | |
| 15-18 | 2.02 | 2.39 | 0.02 | -2.35 | 4.19 | 2.95 | 1.54 | 0.13 |
| | -0.43 | | | | -0.72 | | | |
| 18-21 | 2.80 | 2.30 | 0.66 | -0.99 | 1.48 | 1.81 | -0.63 | -3.07 |
| | -0.62 | | | | -0.42 | | | |

Modified Hill tail estimates are calculated for common, lower and upper tails using the Huisman et alia (2001) weighted least squares regression approach. Standard errors based on linearly interpolated number of tail values for the modified Hill tail estimator are presented in parenthesis for each tail value. Tail estimates are compared to values of 0, 2 and 4 with a critical value of 1.64. 21-24GMT is excluded due to a lack of trading activity for market orders.